\pdfoutput=1 
\documentclass{JINST}

\title{Radiation experience with the CMS pixel detector}

\author{Viktor Veszpremi for the CMS Collaboration$^a$\thanks{Supported by the Janos Bolyai Research Scholarship of the Hungarian Academy of Sciences and the Hungarian Scientific Research Fund under contract number OTKA NK 109703.}\\
\llap{$^a$}Wigner Research Centre for Physics,\\
  1525 Budapest, P.O.Box 49, Hungary\\
E-mail: \email{veszpremi.viktor@wigner.mta.hu}}

\abstract{
The CMS pixel detector is the innermost component of the CMS tracker occupying the region around the centre of CMS, where the LHC beams are crossed, between 4.3~cm and 30~cm in radius and 46.5~cm along the beam axis. It operates in a high-occupancy and high-radiation environment created by particle collisions. Studies of radiation damage effects to the sensors were performed throughout the first running period of the LHC. Leakage current, depletion voltage, pixel readout thresholds, and hit finding efficiencies were monitored as functions of the increasing particle fluence. The methods and results of these measurements will be described together with their implications to detector operation as well as to performance parameters in offline hit reconstruction.
}

\keywords{LHC; CMS; pixel detector, radiation damage}

\begin{document}

\section{Introduction}\label{sec:Introduction}

The CMS tracker~\cite{CMSExperiment} is a silicon detector with a sensitive area of over 200~m$^2$. Its sensors are arranged in concentric cylinders around the LHC beams. Charged particles are induced by the beam collisions that take place in the centre of the detector. The tracker is located inside a 3.8~T homogeneous magnetic field with field lines parallel to the symmetry axis of the silicon barrel. The purpose of the detector is to provide three dimensional position measurements in order to reconstruct the charged particle trajectories. The best tracking performance is achieved in the barrel region within a pseudorapidity range $-0.9<\eta<0.9$. Tracks are used to reconstruct the positions of interaction and decay vertices. The high rates of both ionising and non-ionising particles present a challenge for data-reconstruction especially in the inner layers of the tracker.

\begin{figure}[tbp] 
\centering
\includegraphics[height=3.7cm]{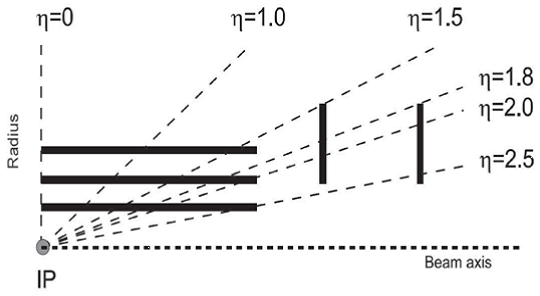}
\caption{Quarter of a slice of the CMS pixel detector by a plane which contains its axis of symmetry. The centre of the detector is at the left-bottom corner of the drawing, in the interaction region. The horizontal axis, to which the detector has a cylindrical symmetry, is parallel to the LHC beams. The vertical axis points along the radius. Various pseudo-rapidity values are shown at the ends of the black dashed lines.
}
\label{fig:PixelRZSlice}
\end{figure}

The pixel detector is the innermost component of the CMS tracker~\cite{CMSPixelDetector}. It was designed utilising a hybrid silicon technology. It comprises three layers in the barrel region at radii of 4.3~cm, 7.2~cm, and 11~cm, respectively (Figure~\ref{fig:PixelRZSlice}). The barrel is closed at the two ends, in the endcap region, with a pair of disks at distances of $\pm$ 34.5~cm and $\pm$ 46.5~cm with respect to
the interaction point, along the beam axis. Assuming a registered hit in the innermost layer (layer~1) of the barrel, this arrangement provides at least three measurement points for tracks originating from the CMS interaction region with a pseudorapidity up to $|\eta|\simeq{2.5}$.

The pixel detector is segmented into 66 million $n+$ pixels of size 100~$\mu$m by 150~$\mu$m implanted into $n$-type bulk with a thickness of 285~$\mu$m and a $p$-type back side. Data sparsification happens in 52~pixel by 80~pixel arrays by readout chips which are bump-bonded on the sensors. Pixels in a chip are arranged into 26 double columns of 160 pixels. These double columns are aligned in the azimuthal direction in the barrel layers and radially in the endcap disks. A total of 15840 chips grouped into modules are read out via 1312 optical links. The serialisation of data transmitted by a group of readout chips into single links is performed by the token bit manager~\cite{PixelROC}.

\begin{figure}[tbp] 
\centering
\includegraphics[width=8cm,height=4.8cm]{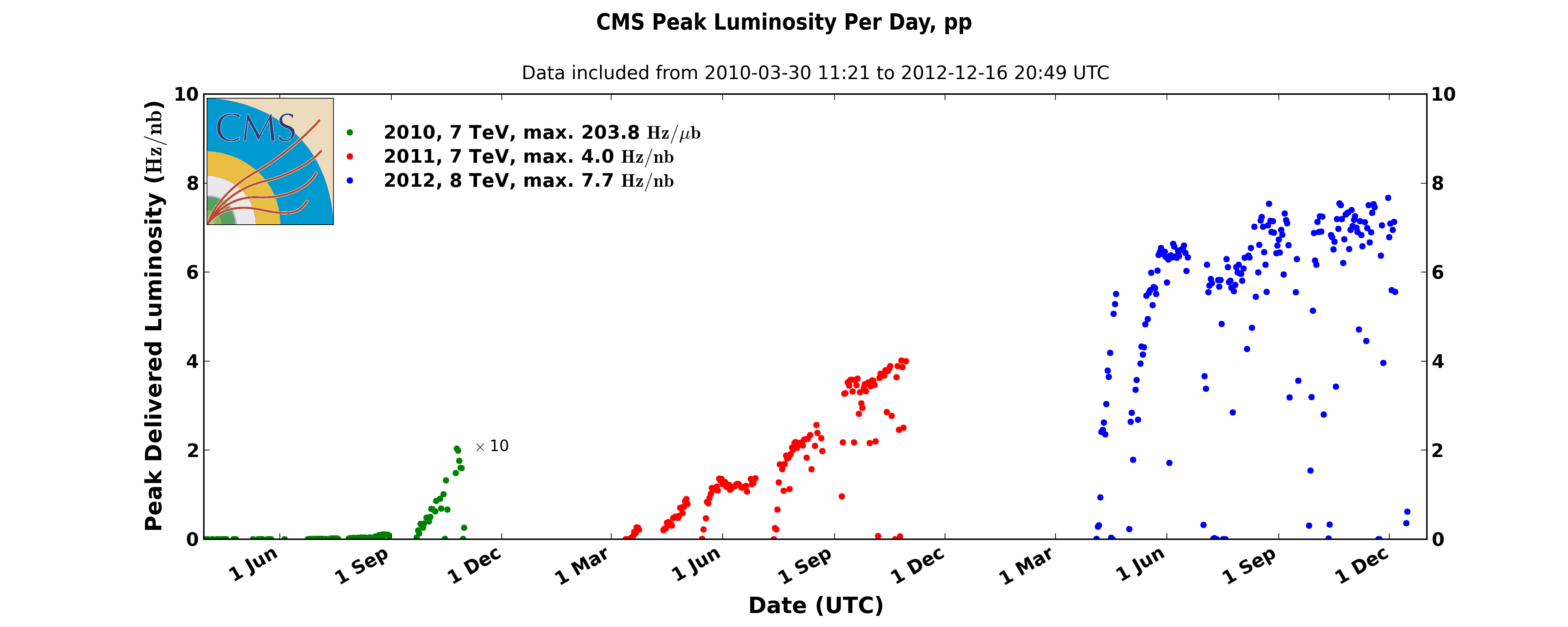}
\includegraphics[width=4.8cm,height=4.8cm]{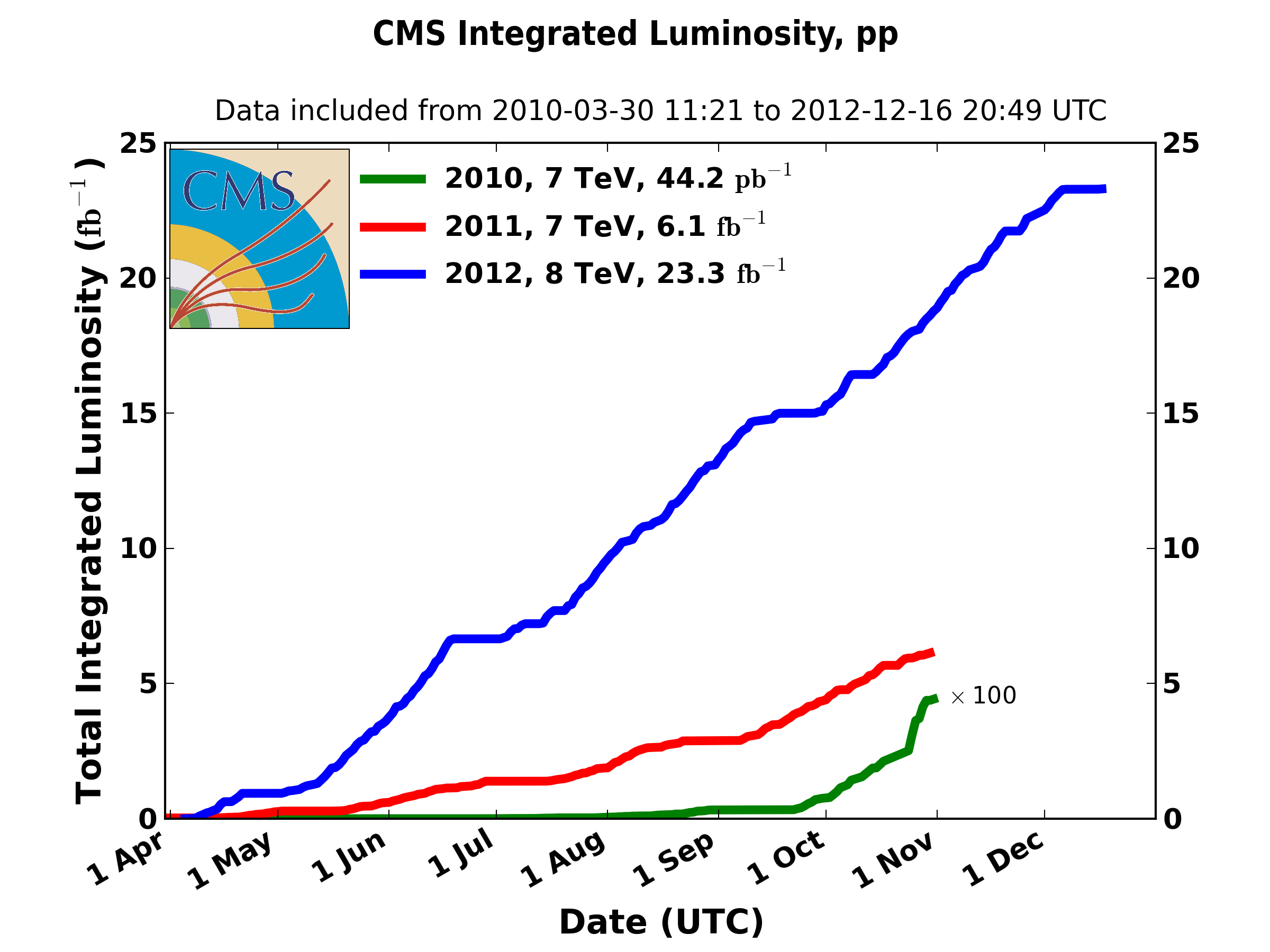}
\caption{LHC beam parameters. Peak (left) and integrated (right) luminosities delivered to CMS during stable beams in p-p 
collisions as function of time. This is shown for 2010 (green), 2011 (red) and 2012 (blue) data-taking~\cite{CMSLumi}.}
\label{fig:LHCBeamParameters}
\end{figure}

The first data-taking period of the LHC~\cite{LHCMachine} between 2010 and 2013 is called Run 1. The instantaneous luminosity of the accelerator gradually improved over the years and reached its peak of 7.7$\times$10$^{33}$cm$^{-2}$s$^{-1}$ in late 2012. During Run 1, the LHC delivered about 6.1~fb$^{-1}$ integrated luminosity of proton-proton collision data at 7~TeV and about 23.3~fb$^{-1}$ at 8~TeV (Figure~ \ref{fig:LHCBeamParameters}). 
The charged track fluence integrated over the 30~fb$^{-1}$ of proton-proton collision data for layer~1, 2, and 3 were 9.3$\times$10$^{13}$ tracks/cm$^2$, 4.2$\times$10$^{13}$ tracks/cm$^2$, and 2.1$\times$10$^{13}$ tracks/cm$^2$, respectively, with about 20\% uncertainty. The 2012 running period was interrupted by two technical stops in mid-June and mid-September. These beam-quiet periods provided the time to re-tune the calibrations of the detectors. The outcome of these calibrations are attested by most of the measurement results presented below. The inner and outer rings of the endcap disks experience a particle rate that is similar to layer~2 and 3 in the barrel. Since the highest particle fluence is measured in layer~1, we will mostly focus on measurement results obtained for the innermost layer.

\section{Radiation effects in the readout electronics}\label{sec:RadiationEffectsInTheRocs}

We have observed both temporary and long-term effects on the readout chips due to irradiation. The short-term effects are always connected to high instantaneous luminosity. This quantity keeps decaying over a fill which reduces the rate of ionising particles. The fluctuation in the particle rate makes consistent measurements and detector studies difficult. Long-term effects, in contrast, are parameterized by integrated luminosity.

\subsection{Temporary effects}\label{sec:ShortTermBeamEffects}

Single event upsets (SEU) are a typical example for temporary effects caused by ionising particles flipping the state of a control register in the readout chips or token bit managers. An SEU may degrade or interrupt data-taking usually resulting in reduced cluster efficiency due to missing clusters. Properties of measured clusters remain unchanged. A reprogramming of all the registers triggered automatically by the readout front-end or manually based on feed-back from data quality monitoring provides remedy to this problem.

\begin{figure}[tbp] 
\centering
\includegraphics[width=4.5cm,height=4.5cm,trim=0 -1.1cm 0 1.1cm]{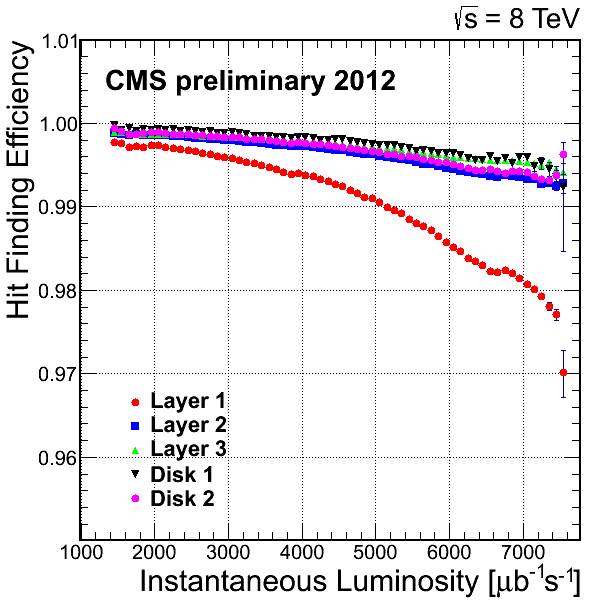}
\includegraphics[width=4.7cm,height=4.7cm]{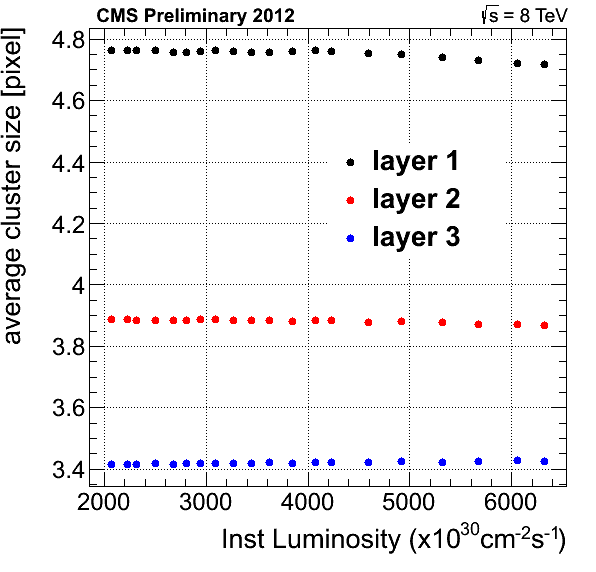}
\caption{Cluster hit efficiency (left) and average cluster size (right) measured in various layers of the pixel detector. Chips experiencing SEUs are removed from the efficiency calculation~\cite{TrackerDPG2013}.}
\label{fig:HitEfficiency}
\end{figure}

Pixel hits are stored in buffers associated to each double column until a decision arrives from the global CMS trigger system whether to read out a collision event. At high particle rates, the buffers may fill up leading to loss of pixel hits in entire columns. In the central region, losing an entire cluster is more probable due to the smaller cluster size. At larger pseudo-rapidity, clusters spread along multiple double columns. These clusters are recorded but often with reduced size and charge. After full track reconstruction, the cluster hit efficiency is measured as the fraction of the particles that are expected to pass through the fiducial regions of the sensors for which corresponding clusters are found within a 500~$\mu$m radius of the expected intersection point. Figure~\ref{fig:HitEfficiency} shows a small reduction of efficiency at high instantaneous luminosity and a slight decrease in the average cluster size. Direct study of the underlying mechanism is not possible, approximate loss rates may be 
inferred from simulation.

The presence of multiple pixel hits in a double column shift the readout threshold in the pixels. Beyond the correlation of their occurrence with particle rate, multiple pixel hits are also produced by Lorentz drift induced charge sharing along columns. Missing pixel hits with charge near their threshold value also leads to reduced cluster size and charge.

\begin{figure}[tbp] 
\centering
\includegraphics[width=4.7cm,height=4.7cm]{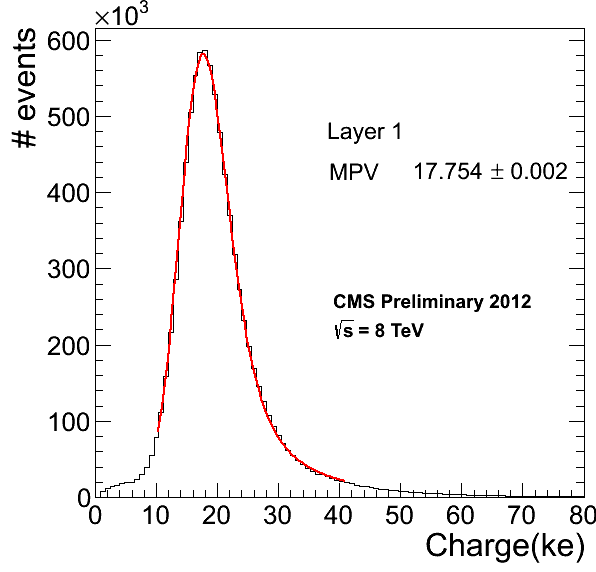}
\includegraphics[width=4.7cm,height=4.7cm,trim=0 0.3cm 0 0]{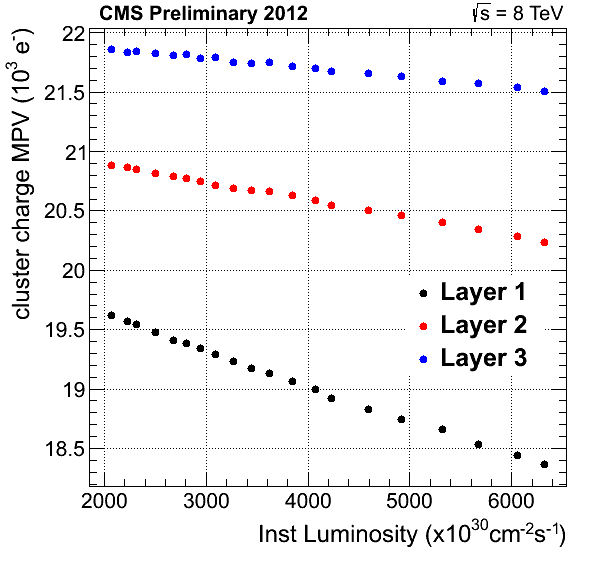}
\caption{Cluster charge distribution normalised with the track incident angle at the end of 2012 (left) and the most probable value of its Landau fit (right) as a function of instantaneous luminosity~\cite{TrackerDPG2013} from a long LHC fill in 2012.}
\label{fig:ClusterParameters}
\end{figure}

A high particle rate increases the power consumption of the readout chips, and therefore raises its operating temperature. The charge gain calibration was observed to be temperature dependent. Figure~\ref{fig:ClusterParameters} displays a variation in the most probable value of the cluster charge as function of the instantaneous luminosity which is more significant than it may be expected based on the variation in cluster size.

A combination of the above effects influences both cluster size and charge, thus drawing solid conclusions about single effects from the variation of these quantities proves to be difficult.

\begin{figure}[tbp] 
\centering
\includegraphics[width=4.7cm,height=4.7cm,trim=0 -0.5cm 0 1.1cm]{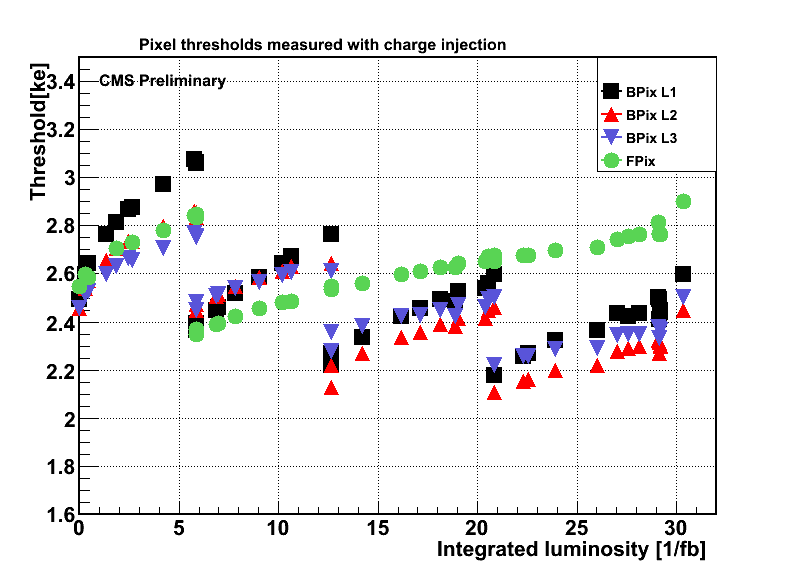}
\includegraphics[width=4.7cm,height=4.7cm]{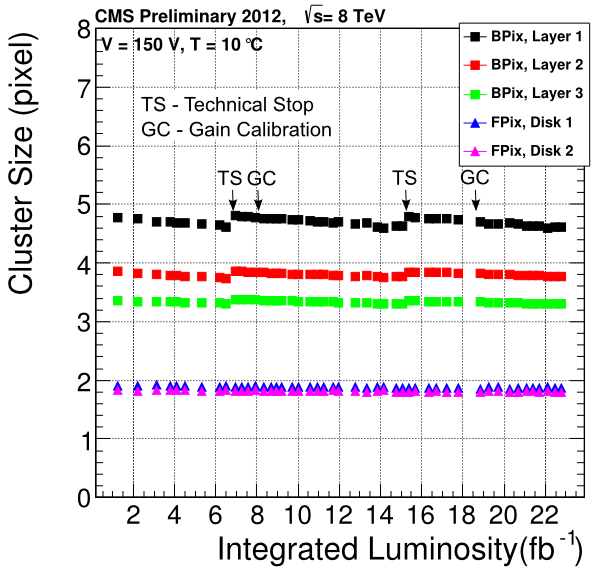}
\includegraphics[width=4.7cm,height=4.7cm]{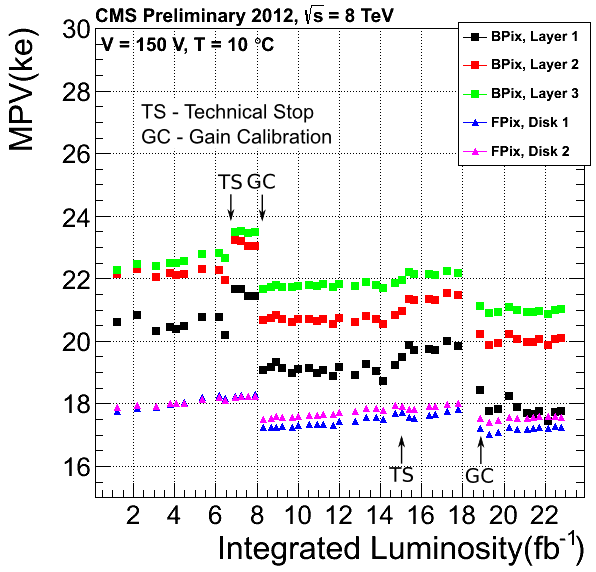}
\caption{Plot of the pixel readout threshold (left) covering the integrated luminosity of Run 1, while data taken in 2012 starts at 6.1~fb$^{-1}$. The other two plots show the evolution of cluster size (middle) and the most probable value of the Landau fit of the cluster charge (right) as function of integrated luminosity in 2012~\cite{TrackerDPG2013}.}
\label{fig:ThresholdVariation}
\end{figure}

\subsection{Cumulative effects}\label{sec:LongTermBeamEffects}

The integrated luminosity delivered by the LHC is used to parameterize long-term, cumulative radiation effects in the readout chips. The radiation dose acquired between 2010 and 2012 was estimated to be about 26~kGy in the innermost 
layer of the pixel detector. 

The left plot in Figure~\ref{fig:ThresholdVariation} shows the evolution of the pixel readout threshold. Thresholds in each layer are measured by injecting test charges into pixel cells in incremental amounts and determining the amount that corresponds to the efficiency turn-on point for half of the pixels. Discontinuities in the plot reveal when threshold readjustment took place: at the beginning of every data-taking year and during technical stops. A lower threshold leads to increased cluster size, which explains the jumps in the second plot (Figure~\ref{fig:ThresholdVariation}), although their magnitude is inconsistent with the fact that later calibrations were observed to achieve lower thresholds. Such an effect may be possible if the unit of the injected test charge shifts. In contrast, the gradual increase in the threshold observed between calibrations should result in diminished cluster size. While such a trend is slightly visible in the barrel layers, it does not appear in the endcap disks (FPix) 
where the thresholds were not readjusted in 2012 at all. The slow increase of the thresholds can, otherwise, be correlated with increased analog current measured at the power supply.

Long term variations are also observed in the cluster charge. The third plot in Figure~\ref{fig:ThresholdVariation} shows the most probable value of a Landau fit on the average cluster charge (see also Figure~\ref{fig:ClusterParameters}). Each sudden increase is due to the increased cluster size after every threshold readjustment. Each sudden drop appears after applying a new gain calibration. Since every threshold calibration is shown to restore the cluster size, the average charge should become automatically readjusted. The overall negative trend, which is observed instead, may imply a change in the gain calibration. The gain calibration associates an ADC value to an injected test charge.

\section{Properties of the irradiated sensors}\label{sec:RadiationEffectsInTheSensors}

\begin{figure}[tbp] 
\centering
\includegraphics[width=7cm,height=4.7cm]{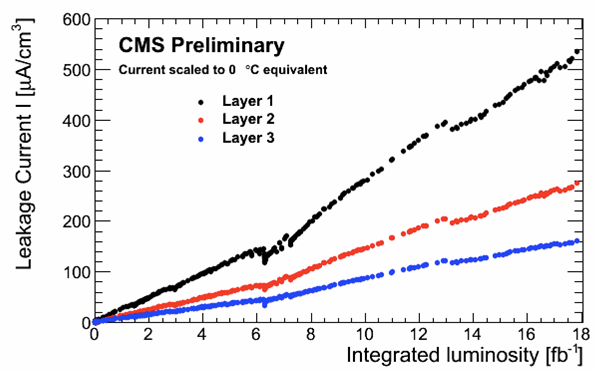}
\caption{Leakage current scaled to $0\,^{\circ}\mathrm{C}$ operational temperature as a function of integrated luminosity until the end of Summer 2012~\cite{LeakageCurrentMeasurement}. }
\label{fig:PixelLeakageCurrent1}
\end{figure}

\begin{figure}[tbp] 
\centering
\includegraphics[width=6.5cm,height=4.9cm]{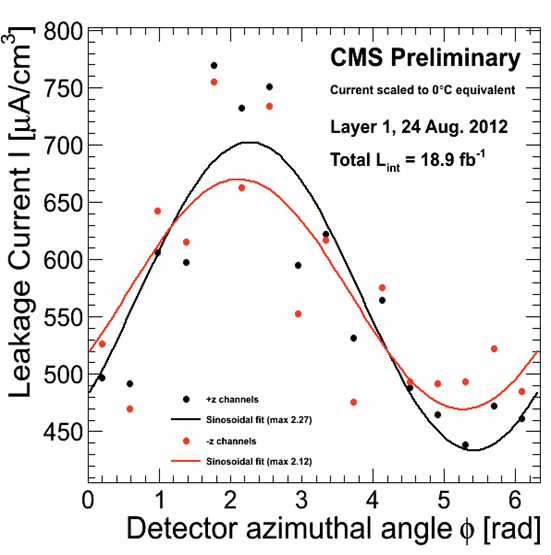}
\includegraphics[width=4.9cm,height=4.9cm]{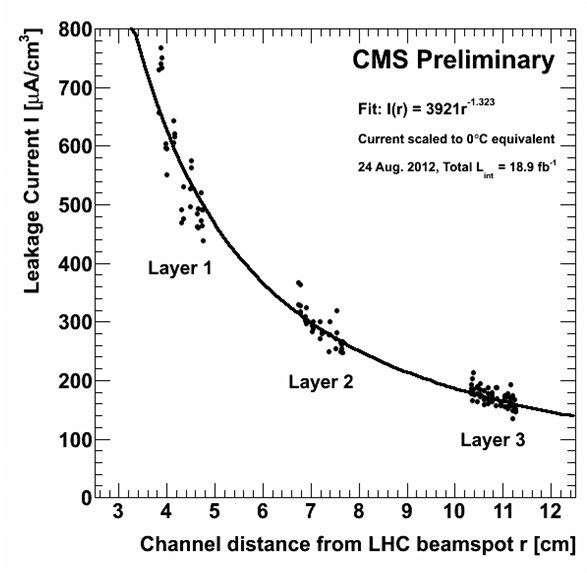}
\caption{Leakage current scaled to $0\,^{\circ}\mathrm{C}$ operational temperature as a function of the azimuthal angle that circles around the symmetry axis of the detector (left) and as a function of the module distance from the LHC interaction region (right). The uneven leakage current in the azimuthal angle is due to an offset in the LHC beam position in the transverse plane. The data points taken in the positive and negative halves of the detector along $z$ are fitted with sinusoidal curves~\cite{LeakageCurrentMeasurement}. }\label{fig:PixelLeakageCurrent2}
\end{figure}

Radiation damage effects induced in the silicon bulk were monitored throughout Run 1. Figure~\ref{fig:PixelLeakageCurrent1} shows the leakage current normalised to unit volume and temperature of $0\,^{\circ}\mathrm{C}$ in the first 18~fb$^{-1}$ of this period.
The leakage current increases linearly with fluence. The drop observed at 6.1~fb$^{-1}$ corresponds to annealing during the shutdown period between 2011 and 2012. In Run~1, the LHC beams were not perfectly centred at the CMS interaction region. The uneven particle rate produced an uneven distribution of leakage current over the azimuthal angle as shown in Figure~\ref{fig:PixelLeakageCurrent2} for layer~1. Each point corresponds to one of the 16 sectors that defines the surface of the barrel. On the top of a sinusoidal shape, a staggered pattern is also observed, which follows the staggered variation of the radial distance of modules in adjacent sectors. This geometry makes it possible to measure the leakage current as function of every module's distance from the interaction region also shown in Figure~\ref{fig:PixelLeakageCurrent2}.

\begin{figure}[tbp] 
\centering
\includegraphics[width=4.8cm,height=4.8cm]{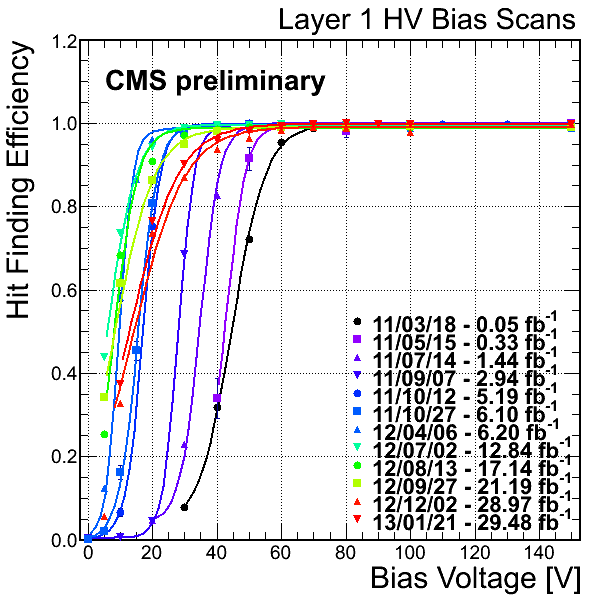}
\includegraphics[width=4.8cm,height=4.8cm]{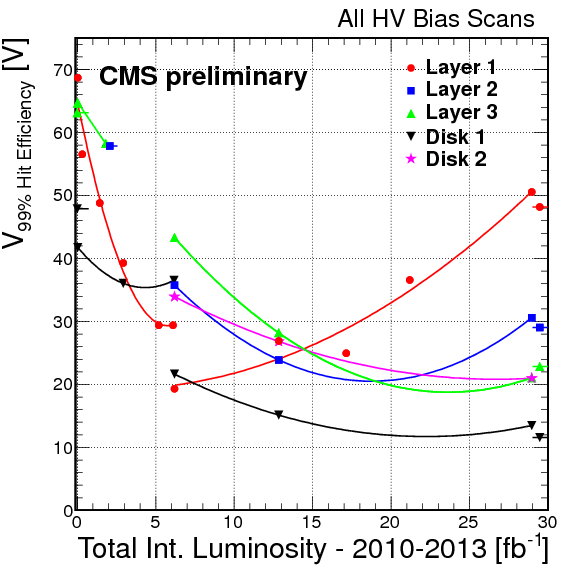}
\caption{Bias scans performed on the pixel detector (left) and the bias voltage corresponding to 99\% hit efficiency (right) as function of the integrated luminosity~\cite{TrackerDPG2013}. Multiple data points acquired in the same year are connected with quadratic fits in the right plot only in order to guide the eye, no underlying model is implied. }
\label{fig:BiasScan}
\end{figure}

The bias voltage applied to the sensors during normal operation is 150~V in the barrel and 300~V in the endcap. In order to measure the full depletion voltage of the sensors, special runs were performed multiple times a year in which the bias voltage was increased in steps from 0 V to the operational voltage. Since a consistent normalisation of the most probable value of the cluster charge has proved to be difficult due to drifting threshold and gain calibration as described in the previous section, the data was evaluated by measuring the cluster hit efficiency. Figure~\ref{fig:BiasScan} shows the efficiency curves as function of the applied high voltage after various amounts of irradiation. The bias voltage needed to reach a depletion depth which corresponds to the full cluster efficiency first decreases, then increases. The second plot shows the evolution of the bias voltage corresponding to 99\% efficiency. Evidence of space charge sign inversion is observed in layer~1. Due to its insensitivity to cluster 
properties, however, this measurement cannot give account on the evolution of the charge collection efficiency when the major factor that describes radiation effects becomes charge trapping. 

\begin{figure}[tbp] 
\centering
\includegraphics[width=4.8cm,height=4.8cm]{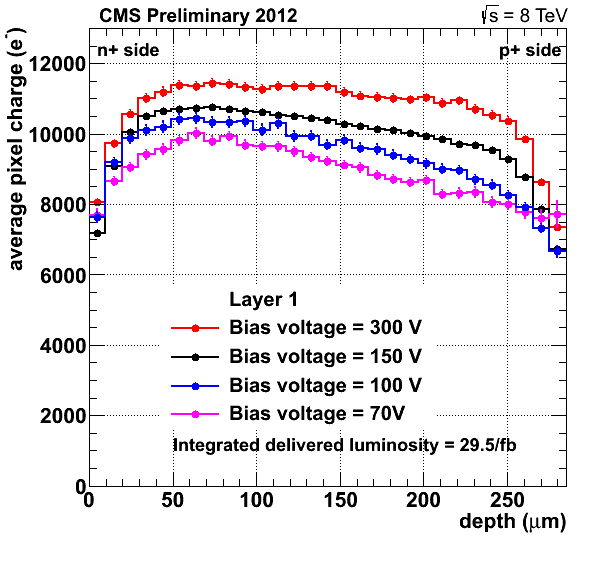}
\includegraphics[width=4.8cm,height=4.8cm]{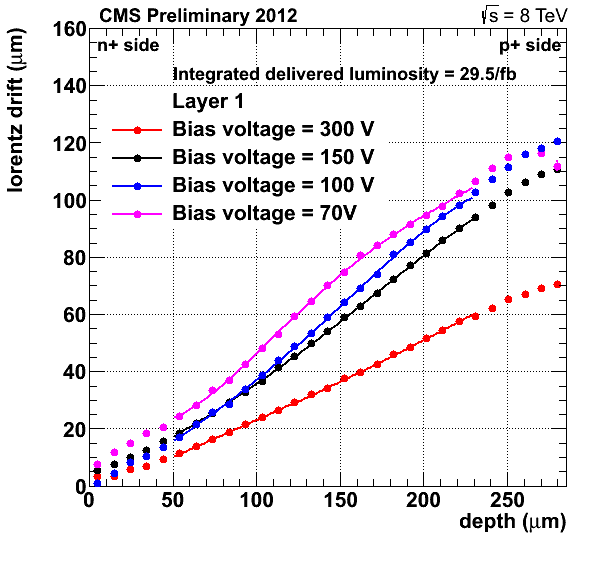}
\caption{Charge collection profile that shows the pixel charge as function of the depth position of the charge deposit (left). Lorentz drift in the transverse plane as function of depth (right)~\cite{TrackerDPG2013}. }
\label{fig:PixelChargeProfile}
\end{figure}

Figure~\ref{fig:PixelChargeProfile} shows the average pixel charge of clusters in layer~1 as function of the depth at which the charge was deposited by the traversing particles. The sensors are read out on their n+ side. The curves show signs of charge trapping which can be mitigated by increasing the bias voltage. The second plot shows the Lorentz drift distance of charges deposited in various depths of the sensor. Lorentz drift induced charge sharing is crucial for good measurement resolution. Charge sharing is reduced upon applying higher bias voltages.

\begin{figure}[tbp] 
\centering
\includegraphics[width=5.2cm,height=5.2cm]{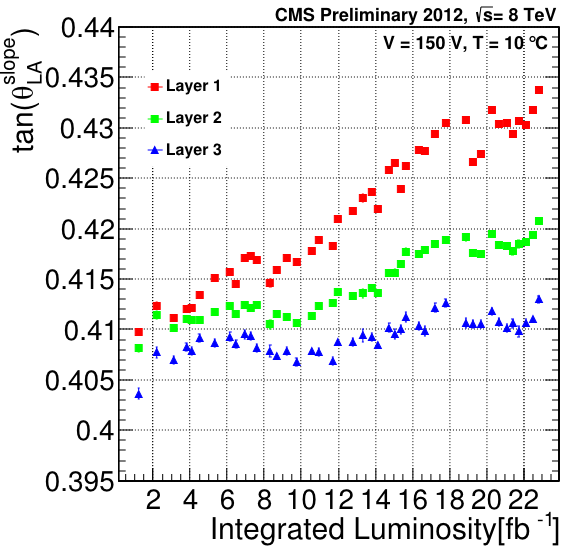}
\includegraphics[width=5.2cm,height=5.2cm]{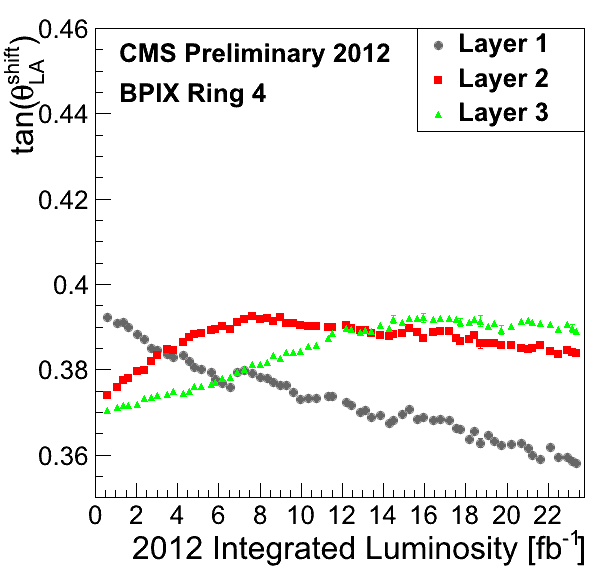}
\caption{Lorentz angle measured near the mid-plane of the sensors (left) and Lorentz shift (right), an apparent movement of the cluster position due to the magnetic field as seen by the tracker alignment method, as function of integrated luminosity in 2012~\cite{TrackerDPG2013,CMSTrackerAlignment2013}. }
\label{fig:LorentzAngles}
\end{figure}

The slope of the Lorentz drift versus depth plot near the sensor mid-plane is called the Lorentz angle. Its evolution with irradiation in 2012 is shown in Figure~\ref{fig:LorentzAngles}. The cluster charge distribution gets wider and more tilted with higher Lorentz angle. However, this is not the only factor that influences the position measurement of a cluster. The second plot shows the shift of the reconstructed cluster position as seen by the tracker alignment algorithm. This shift is also translated into an angle. The reason why the two angles are different is currently not well understood. The value of the Lorentz shift is used as a correction on top of the Lorentz angle in the cluster position measurement algorithm in order to attain the best resolution.

\section{Detector performance}\label{sec:DetectorPerformance}

\begin{figure}[tbp] 
\centering
\includegraphics[width=5cm,height=5cm]{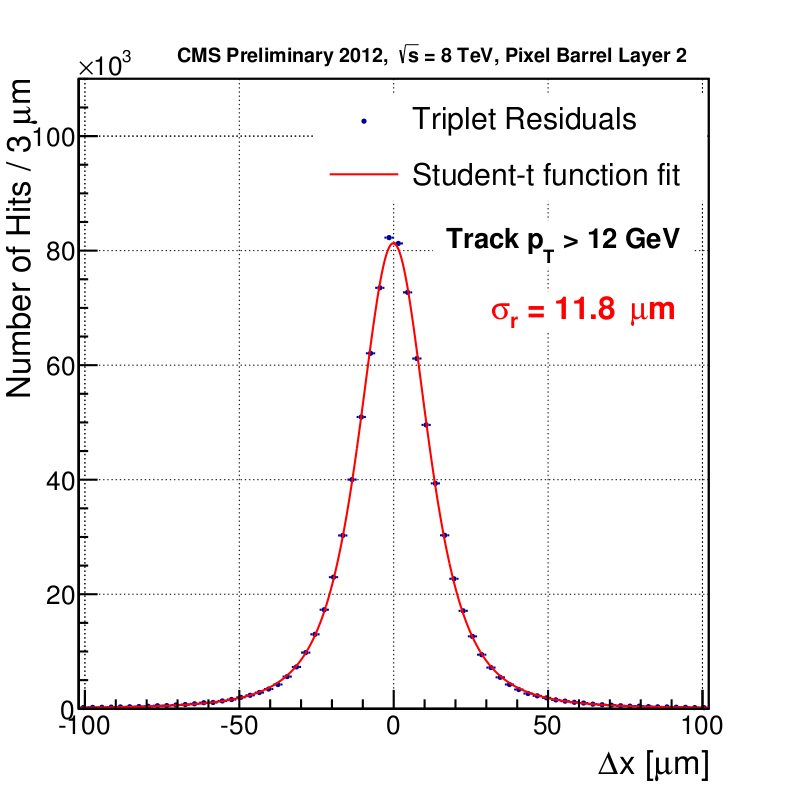}
\includegraphics[width=5cm,height=5cm]{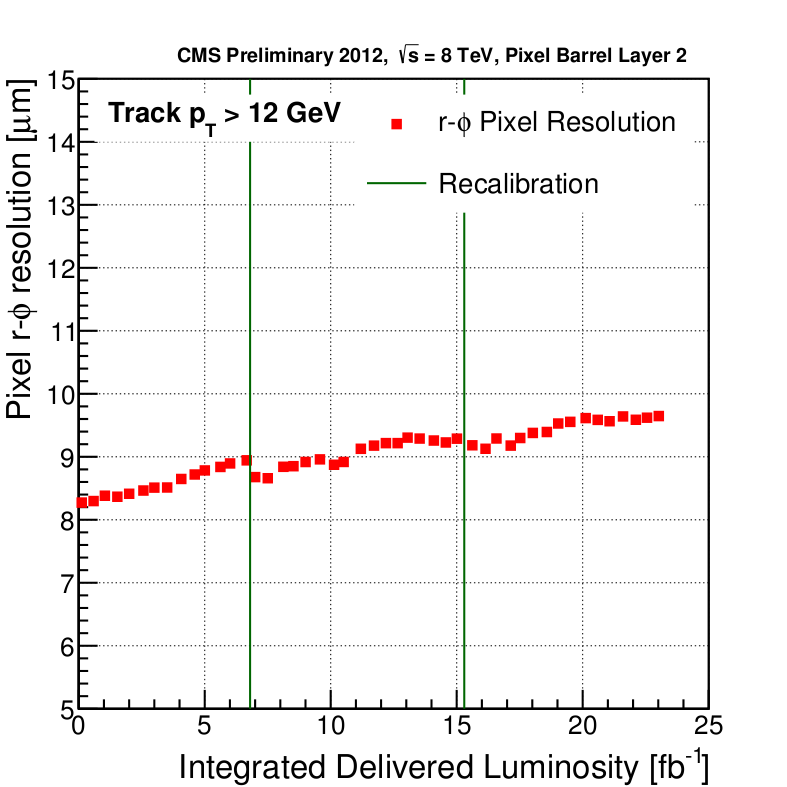}
\caption{Triplet residual distribution for hits in layer~2 measured in the transverse plane (left) and the evolution of the corresponding resolution as a function of delivered integrated luminosity (right)~\cite{TrackerDPG2013}.}
\label{fig:PixelResolutionL2}
\end{figure}

Beside hit efficiency, cluster position resolution is the most important performance parameter of the pixel detector. Even though layer 1 experiences the highest radiation, it also plays the most crucial role in determining track impact parameter resolution, which then translates into primary interaction vertex and secondary decay vertex resolutions. The latter one determines the heavy flavour tagging capabilities of CMS.

The resolution of the barrel pixel is measured by re-fitting hit triplets with their reconstructed tracks whilst keeping the transverse momentum of the tracks and omitting from the fit the hit on the layer under investigation. It is derived from the distribution of the residuals between the expected and measured hit positions, shown in left plot in Figure~\ref{fig:PixelResolutionL2} for layer 2. The measurement in layer~2 is more precise than in layer~1, since hits are used in the fit from radially adjacent layers both below and on the top of layer~2. The resolution is the width of the residual distribution after removing errors which propagate from the fit of the other two hits. Figure~\ref{fig:PixelResolutionL2} on the right shows the resolution of layer~2 as function of integrated luminosity. Its value depends on various factors, such as the precise modelling of the charge sharing, the stability of the gain calibration, and the calibration of the readout thresholds. The re-calibrations of the thresholds 
which happened twice in 2012 are seen as 
slight improvements in the resolution.

\section{Conclusion}\label{sec:Conclusion}

The performance of the CMS pixel detector was excellent during Run~1. At most a couple of per cent efficiency loss was observed at the highest instantaneous luminosities, limited usually to the beginning of the LHC fills. The intrinsic resolution remained close to 10~$\mu$m. Various effects have been identified and monitored that are due to radiation and are expected to affect these performance parameters. However, more significant irradiation effects are expected to appear in Run 2. 
Several improvements are in preparation to take place before and during Run~2. In order to compensate for the accumulated radiation damage, the coolant temperature will be lowered in the detectors. Centring the pixel detector around the LHC beams will result in more uniform distribution of radiation damage along the azimuthal angle. Improved hit reconstruction will take into account the radiation induced changes in the sensors. The simulation of the detector is under development to include the effects of efficiency loss and charge trapping allowing for more detailed studies of the detector performance.


\begin{thebibliography}{9}

\bibitem{CMSExperiment}
S.~Chatrchyan~et~al.~[CMS Collaboration], \emph{The CMS experiment at the CERN LHC}, \jinst{3}{2008}{S08004}.

\bibitem{CMSPixelDetector}
H.~C.~Kaestli~et~al, \emph{CMS barrel pixel detector overview}, Nucl.~Instrum.~Methods~A 582 (2007) 724.

\bibitem{PixelROC}
D.~Kotlinski~et~al, \emph{The control and readout systems of the CMS pixel barrel detector}, Nucl.~Instrum.~Methods~A 565 (2006) 73.

\bibitem{LHCMachine}
L.~Evans~and~P.~Bryant, \emph{LHC machine}, \jinst{3}{2008}{S08001}.

\bibitem{CMSLumi}
\emph{CMS Luminosity - Public Results (2013)},
https://twiki.cern.ch/twiki/bin/view/CMSPublic/LumiPublicResults.

\bibitem{TrackerDPG2013}
\emph{CMS Tracker Detector Performance Results (2013)},
https://twiki.cern.ch/twiki/bin/view/CMSPublic/DPGResultsTRK.

\bibitem{LeakageCurrentMeasurement}
S.~Zenz [CMS Collaboration], \emph{Operational Issues of the Present CMS Pixel Detector}, PoS Vertex2012 (2013) 051.

\bibitem{CMSTrackerAlignment2013}
CMS Collaboration, \emph{2012 CMS Tracker Alignment: Performance Plots}, CERN-CMS-DP-2013-017.


\end{thebibliography}
\end{document}